\documentclass[useAMS,usenatbib]{mn2e}

\usepackage[dvips]{graphicx}
\usepackage[]{txfonts}
\def\simless{\mathbin{\lower 3pt\hbox
{$\rlap{\raise 5pt\hbox{$\char'074$}}\mathchar"7218$}}}   
\def\simmore{\mathbin{\lower 3pt\hbox
{$\rlap{\raise 5pt\hbox{$\char'076$}}\mathchar"7218$}}}   
\def\Msun{{\rm M}_\odot}                                       
\newcommand{\be}{\begin{equation}}
\newcommand{\ee}{\end{equation}}
\title{Fast TeV variability from misaligned minijets in the jet of M87}

\author[Dimitrios Giannios, Dmitri A. Uzdensky and Mitchell C. Begelman]
{Dimitrios Giannios$^{1}$\thanks{E-mail: giannios@astro.princeton.edu
 (DG)}, Dmitri A. Uzdensky$^{1}$ and  Mitchell C. Begelman$^{2,3}$\\
$^{1}$Department of Astrophysical Sciences, Peyton Hall, Princeton
  University, Princeton, NJ 08544, USA\\
$^{2}$Joint Institute for Laboratory Astrophysics, University of Colorado,
  Boulder, CO 80309, USA\\
$^{3}$Department of Astrophysical and Planetary Sciences, University of
  Colorado, Boulder}

\begin{document}
\date{Received / Accepted}
\pagerange{\pageref{firstpage}--\pageref{lastpage}} \pubyear{2009}

\maketitle

\label{firstpage}

\begin{abstract}

The jet of the radio galaxy M87 is misaligned, resulting in a Doppler factor
$\delta\sim 1$ for emission of plasma moving parallel to the jet.
This makes the observed fast TeV flares on timescales of $t_{\rm v}\sim 5R_{\rm g}/c$
harder to understand as emission from the jet. In previous work, we have proposed a
jets-in-a-jet model for the ultra-fast TeV flares with $t_{\rm v}\ll R_{\rm
 g}/c$ seen in Mrk 501 and PKS 2155-304. Here, we show that about half of the
minijets beam their emission outside the jet cone. Minijets emitting off the
jet axis result in rapidly evolving TeV (and maybe lower energy) flares
that can be observed in nearby radio galaxies. The TeV flaring from M87
fits well into this picture, if M87 is a misaligned blazar.

\end{abstract} 
  

\section{Introduction} 
\label{intro}

The jet originating from the nucleus of M87 has been well studied at all wavelengths
thanks to its proximity. Its radio images and modeling of its interaction with its
environment suggest that the jet is misaligned by $\theta\sim
30^{\rm o}$ with respect to our line of sight (Biretta, Zhou \& Owen 1995; 
Bicknell \& Begelman 1996).

Recently several groups (Aharonian et al. 2006; Albert et al. 2008; Acciari et al. 2008) have
reported TeV emission from M87 that shows rapid variability on timescales
of $\sim 1$ day. The fast flaring implies emitting regions of typical length scale of
$l_{\rm em}\simless 5\times R_{\rm g}\delta$, where $R_{\rm g}=GM/c^2$ and $M=3\times 10^9\Msun$
is the mass of the black hole measured by gas dynamics in the
vicinity of the black hole (Macchetto et al. 1997)\footnote{Gebhardt \&
Thomas (2009) claim that the black hole mass is a factor of 2 larger than
previously thought. While in this paper, we keep $M=3\times 10^9\Msun$ as the reference value for
the black hole mass in M87, a larger black hole corresponds to
more extreme variability of the TeV emission that strengthens our main
arguments.} and $\delta$ is the Doppler factor of the emitting material. 
Since the jet from M87 is misaligned by $\sim 30^{\rm o}$, any Doppler factor associated with the
bulk motion of the jet is $\delta\sim 1$, leading to $l_{\rm em}\simless 5 R_{\rm g}$. 

Aharonian et al. (2006) proposed that the TeV emitting region is not associated
with the jet but with the magnetosphere of the black hole (Levinson 2000). 
Alternatively, the TeV emission may originate in the sub-pc jet
while the jet is collimating (Lenain et al. 2008) or in (at least) two
zone models such as a jet with a fast spine and a slower layer
(Tavecchio \& Ghisellini 2008) or a decelerating jet (Georganopoulos, 
Perlman \& Kazanas 2005). The knot HST-1 has also been discussed
as a possible source of the TeV emission (Cheung, Harris \& Stawarz  2007).

The latest (February 2008) observed TeV flaring of M87 was 
caught simultaneously in radio VLBI and X-rays (Acciari et al. 2009).  
At the period of the TeV activity, the radio started to show a gradual rise that 
continued over the succeeding months.  The VLBI map showed new radio blobs 
starting to move outward from within $\sim 100R_{\rm g}$ from the black hole, 
suggesting that the gamma-ray flare comes from very close to the black hole.
While the X-rays from the nucleus were climbing during the flare,
HST-1 was in a low X-ray state, making it unlikely that 
HST-1 is the source of the flares. 

Fast flaring on timescales  $t_{\rm v}\sim R_{\rm g}/c$  
observed in blazars is attributed to large Doppler factors $\delta\gg
1$ of emitting plasma associated with relativistic jets pointing directly at us.
The very fast TeV variability of the blazars Mrk 501 and PKS 2155-304
(Aharonian et al. 2007; Albert et al. 2007)
corresponds to $t_{\rm v}\sim 0.1 R_{\rm g}/c$ and poses strong constraints to any model.
For the TeV radiation to avoid absorption at the source, the emitting
material must move with a bulk Lorentz factor $\Gamma_{\rm em}\simmore 50$ (Begelman, Fabian \& Rees
2008) while resolved patterns
on sub-pc scales are compatible with moderate Lorentz factors
$\Gamma_j< 10$ for the jet (Piner \& Edwards 2004; Giroletti et al. 2004).
In a previous work, we proposed that very rapid TeV variablity observed
in blazars comes from blobs of energetic particles that are moving 
with relativistic speed {\it relative to} the mean flow of the jet (Giannios, Uzdensky \& Begelman 2009). 
We associate these minijets with the material outflowing from reconnection
events in a Poynting-flux dominated jet. Jets with typical
$\Gamma_{\rm j}\simless 10$ (as indicated by proper motions) would thus contain minijets that
move with $\Gamma_{\rm em}\sim 100\gg\Gamma_j$ relative to the observer, as
needed for TeV radiation to escape the source and power flares with $t_v\ll
R_{\rm g}/c$.

A natural consequence of the jets-in-a-jet model is that, while the 
emission of about half of the minijets is beamed at an angle 
$\theta\simless 1/\Gamma_{\rm j}$ with respect to the jet axis, 
the rest emit outside the emission cone of the jet.
In this paper, we explore the emission from these ``misaligned'' minijets.
We show that the probability that a minijet is {\it observed} to
emit at an angle $\theta$ depends rather weakly on $\theta$. 
We calculate the energetics and timescales of emission from minijets pointing off 
the jet axis. Off-axis minijets are bright enough to be observed in nearby
radio galaxies. Finally, we apply the jets-in-a-jet model to the fast
TeV flaring observed in M87.

\section{Jets in a jet}

Our main idea of ``jets-in-a-jet'' is that dissipation of magnetic energy 
in the jet (e.g., the result of jet instabilities) leads to minijets of energetic particles that
move relativistically {\it within} the jet. Emission from 
the minijet results in the observed short-timescale flaring.
Depending on the direction of motion of the minijet (in the jet rest frame),
a minijet emits within the cone of the jet or outside it. 
In jets moving toward our line of sight (blazars), emission from minijets
results in powerful, rapidly evolving flares  while in  misaligned 
jets (e.g., in radio galaxies), minijets lead to weaker but still rather 
rapid flares.   

We believe that jets-in-a-jet may be expected in a Poynting-flux dominated
flow. They can form in reconnecting regions of strongly magnetized plasma.
The magnetic reconnection may result from the nonlinear evolution of kink instabilities
in the jet (Eichler 1993; Begelman 1998;  Appl, Lery \& Baty 2000; Giannios \& Spruit 2006; 
Moll, Spruit \& Obergaulinger 2008; Moll 2009; but See McKinney \& Blandford 2009) or reversals of the
magnetic field polarity in the inner disk/black hole magnetosphere (Giannios et al. 2009).
Interestingly, recent evidence for such reversals come from the observed ``flip'' in 
the gradient of the Faraday rotation across the jet of B1803+784 between
the 2000 and 2002 observations of the source (Mahmud, Gabuzda \& Bezrukovs
2009). A similar scenario of relativistic motions
within the jet has also been explored in the context of gamma-ray bursts (GRBs).  Minijets
(caused by relativistic turbulence or magnetic reconnection) may be responsible for
the observed variability of the GRB emission
(Blandford 2002; Lyutikov 2006a,b; Narayan \& Kumar 2008; Lazar, Nakar \& Piran 2009).

Consider a jet that moves radially with bulk $\Gamma_{\rm j}$, in which
a minijet develops with a Lorentz factor
$\Gamma_{\rm co}$ at an angle $\theta'$ with respect to the 
radial direction (both measured in the jet rest frame). 
All primed/tilded quantities are measured in the rest frame of the jet/minijet respectively. 
In the lab frame the minijet moves with
\be
\Gamma_{\rm em}=\Gamma_{\rm j} \Gamma_{\rm co}(1+v_jv_{\rm co}\cos \theta')
\label{gamma}
\ee 
and at an angle 
\be
\tan\theta=\frac{v_{\rm co}\sin\theta'}{\Gamma_{\rm j}(v_{\rm
    co}\cos\theta'+v_{\rm j})}
\label{angle}
\ee
with respect to the radial direction. 
For $\theta'\sim \pi/2$ and $v_{\rm co}, v_{\rm j}\simeq 1$, expressions
(1) and (2) give $\Gamma_{\rm em}\sim \Gamma_{\rm j} \Gamma_{\rm co}$ and 
$\theta\sim 1/\Gamma_{\rm j}$. Minijets moving at $\theta>1/\Gamma_{\rm j}$
have more moderate bulk Lorentz factor in the lab frame but, in many cases, larger
than that of the bulk of the jet $\Gamma_{\rm j}$ (see next section).
 
From eq.~(2), one can immediately see that for $\theta'<\pi/2$ the emission
takes place at $\theta<1/\Gamma_{\rm j}$, i.e., within the emission cone of 
the jet and vice versa. 
In the magnetic reconnection picture, every dissipation event results
in a pair of mini jets that leave the reconnection region at opposite 
directions $\theta'$ and $\pi-\theta'$. {\it Since  the
distribution of $\theta'$ of the minijets is symmetric around $\theta'=\pi/2$, for
every minijet that is beamed within the jet $1/\Gamma_{\rm j}$ emission cone there
is another minijet that points outside.}

\subsection{Distribution of emitting angles of minijets with respect to the
  jet axis}

In this section, we explore how the angular dependence of the emission 
from the minijets is connected to their assumed distribution of directions
in the jet rest frame. We find that the 
probability that the observer is located within the emission cone of a minijet
is rather insensitive to the inclination of the jet with respect to the
observer.    

We consider a relativistic conical jet with opening angle 
$\theta_{\rm j}\sim 1/\Gamma_{\rm j}$. Emission associated 
with the bulk motion of the jet is beamed 
at the observer when the latter is located at
an angle $\theta\simless 1/\Gamma_j$ with respect to the jet axis.
An ``off-axis observer'' is one for which $\theta>1/\Gamma_j$. 
We measure how ``off-axis'' the observer is located by the 
parameter $\alpha$ defined as:
\be
\theta\equiv \frac{\alpha}{\Gamma_j}.
\ee

If an observer is located at an angle $\theta$ with respect
to the jet axis, how likely is it that the  emission from a minijet 
takes place within the observer's cone?
As discussed above, in the lab frame, the majority of minijets point at small
angle $\theta$ or corresponding $\alpha\simless$ a few. On the other hand, 
minijets pointing more off-axis have lower bulk Lorentz factors and therefore,
wider beaming angles as we will see. The net result is that there is almost uniform 
probability that a minijet is observed at any angle $\theta$.

To quantify this statement, we consider several different cases for the    
distribution $P(\Omega')d\Omega'$ of the
angle $\theta'$ of the minijets in the jet rest frame.
We consider distributions that are symmetric around $\theta'=\pi/2$,
as expected from the formation of pairs of counter-streaming minijets that result
from magnetic reconnection. We consider 3 cases: $P(\Omega')\propto \sin^n\theta'$,
for $n=0,1,2$. The $n=0$ case corresponds to the isotropic distribution.
For $n=1,2$ the distributions increasingly peak around $\theta'=\pi/2$. 
We think it is very plausible that 
the reconnecting current sheets are preferentially oriented 
perpendicular to the jet direction (e.g., a sequence of toroidal 
fields with reversing radial currents) leading to the ejection of
minijets preferably at $\theta'\sim \pi/2$. 

We assemble a large number of minijets with $\theta'$ drawn from the 
above distributions and calculate the distribution $P(\alpha)d(\alpha)$
of emission angle $\alpha$ using the expression (2). For the results shown 
in Fig.~1, we have set $\Gamma_{\rm j}=\Gamma_{\rm co}=10$. Very similar 
angular distributions are found for other values of $\Gamma_j,\Gamma_{\rm co}\gg 1$.  
The distribution of direction of the minijets
peaks at $\alpha\sim 1$ (i.e., at angle $\theta\sim 1/\Gamma_{\rm j}$ ). 
For increasing $n$, the $0.5\simless \alpha\simless 2$ angles
become more likely. This is expected since the number of 
minijets moving at $\theta'\sim \pi/2$ (which results in
$\alpha\sim 1$) increases with $n$. 
Furthermore, by definition of the symmetry of the distribution
around $\theta'=\pi/2$, about half of the jets point at $\alpha>1$,
independently of $n$.   

\begin{figure}
\resizebox{\hsize}{!}{\includegraphics[angle=270]{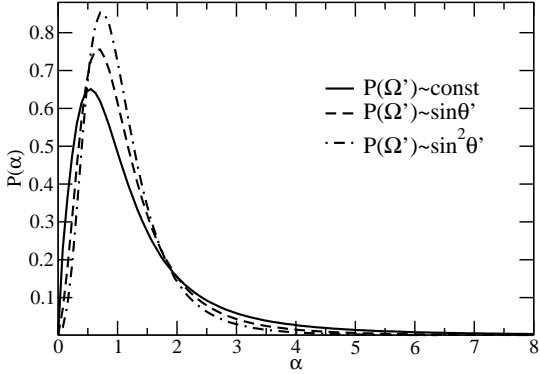}}
\caption[] {Distribution $P(\alpha){\rm d}\alpha$ of direction of minijets
in the lab frame (where $\alpha=\theta\Gamma_j$) for different distributions 
$P(\Omega'){\rm d}\Omega'$ of direction of the minijets (in the jet rest frame).    
\label{fig1}}
\end{figure}

To calculate the probability distribution that a blob is 
{\it observed} at an angle $\theta\equiv \alpha/\Gamma_{\rm j}$ one has to take into
account that the emission of a blob is beamed into a narrow cone
of $\Omega_{\rm em}\propto 1/\Gamma_{\rm em}^2$. In Fig.~2, we show the distribution
$P(\alpha)d\Omega/\Gamma_{\rm em}^2$ which we refer as the ``observed angular
distribution'' $P_{\rm obs}(\alpha)d\Omega$ of the minijets.
 
From Fig.~2, it is clear that for $n=0$ the observed angular
distribution of the minijets is almost isotropic. For $n=1,2$ it
slightly favors $0.5\simless \alpha\simless 3$. The reason for $P_{\rm obs}$
being much flatter than $P$ is that, although most minijets
(per unit of solid angle) point within the jet cone, their emission is beamed
within a very narrow cone. Minijets pointing to larger angles are more broadly
beamed. As a result, the probability that a blob is observed at an angle $\theta$ depends weakly 
on $\theta$.

Out of $N$ minijets forming in the jet, approximately
$N/\Gamma_{\rm co}^2$ will be beamed at an observer located at any angle
with respect to the jet axis. The observed duration and intensity of these flares
will, however, depend on the observer's angle with respect to the jet axis, with 
on-axis observers seeing far more powerful flares. We quantify this point 
in the next section.   

\begin{figure}
\resizebox{\hsize}{!}{\includegraphics[angle=270]{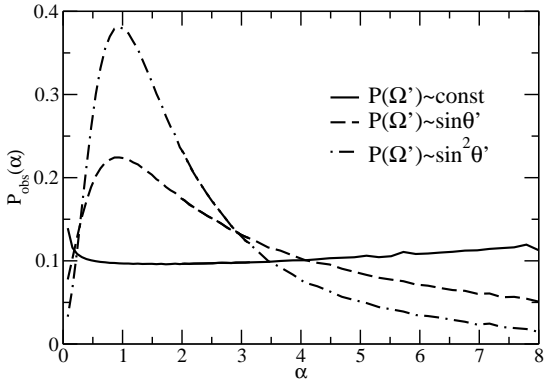}}
\caption[] {Probability (not normalized) $P_{\rm obs}(\alpha)d\Omega\propto
P(\alpha)d(\Omega)/\Gamma_{\rm em}^2$ that a minijet is observed at angle
$\alpha$ for different distributions $P(\Omega'){\rm d}\Omega'$
of direction of the minijets (in the jet rest frame).     
\label{fig2}}
\end{figure}

\subsection{Dependence of observed properties of the minijets on the inclination
to the jet axis}
 
As we have shown in the previous section, the probability that a minijet emits within the observer's
line of sight depends only weakly on the inclination of the observer with respect
to the jet axis. However, the detectability of emission of a minijet
depends on the luminosity of the source. The bolometric
luminosity of a minijet is related to the total
radiated energy of the minijet $E$, the solid angle $\Omega_{\rm em}$ over which the emission
takes place, and the observed duration $\delta t_{\rm obs}$ of the emission:
$L\simeq E/\Omega_{\rm em} \delta t_{\rm obs}$. In the following we discuss 
how  $E$, $\Omega_{\rm em}$ and  $\delta t_{\rm obs}$ depend on the
inclination of the minijet. 

The expressions (1) and (2) give the velocity of the minijet in the lab frame
once the comoving angle $\theta'$ is specified. Here, we derive
much simpler, approximate expressions for the velocity and direction of motion
of the minijet that points at $\theta>1/\Gamma_j$ (off the jet axis).
In the following we express the various quantities as functions of the
inclination $\alpha$.  
In the limit where $\Gamma_{\rm j}$, $\Gamma_{\rm co}\gg 1$
and $1\simless \alpha \simless \Gamma_{\rm j}, \Gamma_{\rm co}$,
eq.~(1) for the Lorentz factor of the blob becomes
\be
\Gamma_{\rm em,app}\simeq \frac{2\Gamma_{\rm j}\Gamma_{\rm co}}{\alpha^2}.
\label{gamma_app}
\ee 
and eq.~(2)
\be
\theta=\frac{2}{\Gamma_{\rm j}(\pi-\theta')},\quad {\rm or} \quad \alpha=\frac{2}{\pi-\theta'}.
\ee

In Fig.~3, we show the fractional error of expression (\ref{gamma_app})
with respect to the exact one [eq.~(1)] as function of inclination $\alpha$
and for different values of $\Gamma_{\rm j}$, $\Gamma_{\rm co}$. 
For $\Gamma_{\rm j}$, $\Gamma_{\rm co}\simmore 5$, the approximate
expression is rather accurate provided that $\alpha\simmore 2$.
In the analytical estimates that follow we use the approximate 
expression (\ref{gamma_app}). 

\begin{figure}
\resizebox{\hsize}{!}{\includegraphics[angle=270]{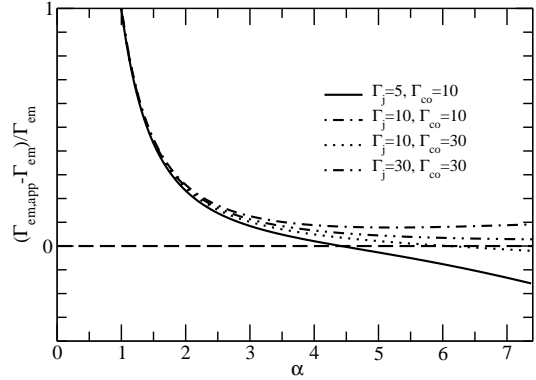}}
\caption[] {Fractional accuracy of approximate expression (\ref{gamma_app})
of the bulk Lorentz factor of the minijet compared to the exact 
expressions (1) and (2) as function of inclination $\alpha$. 
The different curves correspond to different
values of $\Gamma_{\rm j}$ and $\Gamma_{\rm co}$. For $\alpha\simmore 2$,
the approximate expression gives the correct answer within better than 
$20\%$.    
\label{fig3}}
\end{figure}

From eq.~(4), we find that a minijet moves relativistically towards
the observer for a wide range of inclinations $\alpha<(2\Gamma_{\rm j}\Gamma_{\rm co})^{1/2}$.
This leads to the possibility of emitters with high Doppler factor $\delta\gg
1$ even in cases where the jet is misaligned by large angles $\theta\simmore
30^{\rm o}$.

We assume that all the minijets are identical in a frame comoving with the jet
and differ only in their orientation $\theta'$.
Suppose that the energy of the minijets is $\tilde{E}$ (in the minijet rest frame).
When $\theta\simless 1/\Gamma_{\rm j}$ (i.e., $\alpha\simless 1$), a blob has 
lab-frame energy $E_{\rm on}=\Gamma_{\rm em}\tilde{E}\sim 
\Gamma_{\rm j}\Gamma_{\rm co} \tilde{E}$ while for $\theta>1/\Gamma_{\rm j}$
the energy of the minijet is
$E_{\rm off}\sim 2\Gamma_{\rm j}\Gamma_{\rm co} \tilde{E}/\alpha^2\sim 2E_{\rm on}/\alpha^2$.
Since the emission of the minijet takes place within a solid angle 
$\Omega_{\rm em}\sim 1/\Gamma_{\rm em}^2$, the emission from minijets that point off the
jet axis is less beamed than that of on-axis ones: $\Omega_{\rm off}\sim
\alpha^4\Omega_{\rm on}/4$.

The observed timescale of emission from a minijet depends on the size of the blob,
its bulk Lorentz factor, and the radiative mechanism or the duration of the
reconnection event. Identical blobs (i.e., of the same size, density, magnetic
field strength) that emit through the synchrotron and 
synchrotron-self-Compton (SSC) mechanisms have the same emission
timescale $\tilde{\delta t}$ in the rest frame of the minijet. 
The observed duration of the blob emission is $\delta t_{\rm obs}=\tilde{\delta
t}/\Gamma_{\rm em}$. Thus, the duration of the flare for an off axis observer 
is\footnote{If the observed timescale of emission is determined by the duration of the
reconnection event, the duration of the flare for an off-axis observer can
 similarly shown to be $\delta t_{\rm off}\sim \alpha^2\delta t_{\rm on}/2$.} 
$\delta t_{\rm off}\sim \alpha^2\delta t_{\rm on}/2$. This expression shows 
that the timescale of emission becomes longer when the jet is viewed from larger inclination. 
On the other hand, the dependence on $\alpha$ is rather moderate. 
If minijets result in the very rapidly evolving flares of the blazars  
Mrk 501 and PKS 2155-304 with $\delta t_{\rm on}\sim 0.1 R_{\rm g}/c$ (Giannios et
al. 2009), misaligned jets with $\alpha\sim$ a few can also lead to rather 
short variability timescales $\delta t_{\rm off}\sim R_{\rm g}/c$.

Combining the above estimates,
if the minijets emit through synchrotron and SSC, the bolometric luminosity
during the flare of an off-axis blob is
\be
L_{\rm off}\sim \frac{E}{\Omega_{\rm em} \delta t_{\rm obs}} \sim\frac{16L_{\rm on}}{\alpha^8}.
\label{Lblob}
\ee    
Although the $\alpha^8$ dependence is very steep, making 
flares from misaligned jets rather weak, it is still possible to observe 
them from nearby sources. 

The gamma-ray emission from the minijets may not be dominated by
SSC, but by inverse Compton scattering of an external source of seed
photons, provided that a powerful enough source exists. The accretion disk 
and the broad line region have been frequently considered in this
respect (Dermer, Schlickeiser \& Mastichiadis (1992); Sikora, Begelman \& Rees
(1994)). In the jets-in-a-jet model discussed here, 
soft (synchrotron) emission emitted at a different location in 
the jet can also be considered as an external photon field {\it from the point of
view} of the minijet. Such a photon field, approximately isotropic in the 
rest frame of the jet, may be connected to the process responsible for the 
quiescent jet emission or may come from the averaged synchrotron emission of all
the minijets forming during the flaring activity of the jet.
One can show that the expression (6) for the dependence of the blob's luminosity  
on its inclination to the jet axis holds for the external inverse Compton
mechanism, {\it provided that} the photon field is 
approximately isotropic in the rest frame of the jet.
On the other hand, if there is a significant source of
soft radiation coming from the base of the jet (e.g., the accretion
disk), the scaling (6) fails.  

As we shall see in the next section, the observed soft emission coming from the 
nucleus of M87 (associated with the accretion disk and/or the inner jet) 
provides plenty of seed photons to be inverse Compton scattered in the
minijet. External inverse Compton appears to be an important mechanism
for the gamma-ray emission from this source.    

\section{Application to the flaring TeV emission from M87}

The nucleus of M87 contains a $M\simeq 3\times 10^9 \Msun$ 
black hole (corresponding to light-crossing time of $R_{\rm g}/c\sim 1.5\times 10^{4}$ sec)
measured from gas kinematics on scales of 10's of parsecs
(Macchetto et al. 1997). The M87 jet, believed to originate from the black hole, 
has been extensively observed from radio to TeV and from sub-pc to
kpc scales. 

Modeling of the kpc-scale interaction of the jet with the external medium
constrains the properties of the jet. Though rather model dependent, 
the jet inclination is estimated to be $\theta\sim 30^{\rm o}$ with respect 
to the line of sight, the bulk Lorentz factor of the jet $\Gamma_{\rm j}\sim 5$ 
and the kinetic luminosity $L_{\rm j,true}\sim 10^{43}$ erg/sec (Bicknell \& Begelman 1996;
Reynolds et al. 1996, Owen et al. 2000; Stawarz et al. 2006;
Bromberg \& Levinson 2009). 

The nucleus of M87 and the knot HST-1 (about 60 pc away from the nucleus) 
are active with luminosity of order of $\sim 10^{41}$ erg/sec each
over a wide range of wavelengths (from IR to gamma-rays; see, e.g.,
Sparks, Biretta \& Macchetto 1996; Perlman et al. 2001). 
Both the nucleus and the knot HST-1 have shown variability in X-rays
and optical on timescales ranging from months to years (see, e.g., Harris et
al. 2009) indicating rather compact emitting regions. During 2005, M87 
was luminous in the TeV band with isotropic $L_{\rm TeV}\sim 3\times 10^{40}$ 
erg/sec and exhibited flares evolving on $\sim 2$-day timescales (Aharonian et al. 2006). 
The poorer angular resolution in TeV energies
allows for both the nucleus and HST-1 as possible sources of the
high-energy emission. During the 2005 TeV flaring, HST-1 was at its peak of 
its optical and X-ray activity (exceeding the nuclear emission), making 
the knot a promising candidate for the origin of the flares (Cheung et al. 
2007). In February 2008, TeV flaring on $\sim 1$ day timescales 
was observed from M87, this time not associated with high activity from the 
knot HST-1. The correlation of TeV and X-ray fluxes on month to year 
timescales in the nucleus favors the nuclear region as the source of the TeV
flares (Acciari et al. 2008; Harris et al. 2009). Even more convincingly,
the 2008 TeV flaring was followed by new radio blobs 
moving outward from only $\sim 100R_{\rm g}$ from the black hole, suggesting 
that the gamma-ray flare comes from very close to the black hole (Accuari
et al. 2009).
 
If the jet of M87 is misaligned (pointing at an angle $\theta\sim 30^{\rm o}$ 
away from the observer), the associated 
Doppler factor is $\delta=1/\Gamma_{\rm j}(1-v_j\cos \theta)\sim 1$. In this
case relativistic effects do not contribute to reduce variability timescales.
The physical size of the TeV source is very compact,
$l_{\rm em}\simless ct_{\rm v}\sim 5 R_{\rm g}$. This points to the possibility that the emission 
takes place in the vicinity of the black hole. TeV emission coming directly 
from the magnetosphere of the black hole is proposed by Neronov \& Aharonian 
(2007; see also Levinson 2000). A potential problem with this
interpretation of the TeV emission is that
infrared and optical radiation originating from the accretion disk may be
 strong enough to absorb TeV photons. The 
modeling of the emission of the accretion flow and fits to
observations of the nuclear emission indicate that TeV
emission is absorbed if produced at $R\simless 20R_{\rm g}$
(Li et al. 2009). During the flaring, the high-energy emission 
extends up to at least $\sim 10$ TeV  with the TeV spectrum 
well described by a power-law model with a photon-number index of $\Gamma\sim
-2.3$ (Aharonian et al. 2006; Albert et al. 2008). 
There is no evidence for absorption of the TeV emission, suggesting that the
emission takes place at $R_{\rm em}\simmore 20 R_{\rm g}$.   

It has also been proposed that the TeV flares come directly from the inner (pc-scale) regions of
a misaligned jet. These models are more complex than single-zone SSC models and either invoke
jet deceleration on sub-pc scales (Georganopoulos et al. 2005) or a multi-zone
configuration with a fast spine and slower outer layer (Tavecchio \& Ghisellini 2008). 
One weakness of these models is that they tend to produce steeper TeV spectra
than those observed, because of absorption by the dense soft (synchrotron)
photon field along the line of sight.
 
Our jets-in-a-jet picture provides a way out of these difficulties.
While the jet is misaligned with respect to the observer, some minijets
point outside the jet emission cone and in the direction to the observer. 
These ``off-axis'' minijets move relativistically toward the observer
and can be compact, resulting in TeV flares with short variability
timescales. Furthermore, the relativistic beaming toward the observer
allows the TeV emission to escape the production site without
being significantly absorbed.

\subsection{Jets in the jet of M87}

For more quantitative estimates we consider a jet with (isotropic) luminosity $L_{\rm j,iso}$  
that moves with bulk $\Gamma_{\rm j}$. The jet is assumed to be strongly
magnetized with a Poynting-to-kinetic flux ratio (magnetization) $\sigma\gg 1$. 
As reference values, we use $\Gamma_{\rm j}=5$
and $\sigma=100$. The isotropic equivalent jet luminosity  
is estimated to be $L_{\rm j, iso}= 4\Gamma_{\rm j}^2L_{\rm j,true}\sim 10^{45}$ erg/s. The
energy density and  magnetic field strength in the jet as function of radius
are (e.g., Giannios et al. 2009):
\be
e'_{\rm j}=L_{\rm j, iso }/4\pi r^2c\Gamma_{\rm j}^2=0.05 L_{\rm
  j,45}r_2^{-2}\Gamma_{\rm j,5}^{-2}\quad \rm{erg/cm^3},
\ee
and
\be
B'_{\rm j}=\sqrt{4\pi e'_{\rm j}}=0.8  L_{\rm j,45}^{1/2}r_2^{-1}
\Gamma_{\rm j,5}^{-1}\quad {\rm Gauss},\ee
where  $A=10^xA_x$ and the spherical radius is $R=r\cdot R_{\rm g}$ with 
$R_{\rm g}=4.5\times 10^{14}$ cm, corresponding to the gravitational radius of a
black hole of $3\times 10^9\Msun$.  

We assume that a fraction of the magnetic energy of the jet 
is occasionally dissipated through reconnection. 
Our picture for relativistic reconnection is the relativistic generalization of 
Petschek-type reconnection worked out by Lyubarsky (2005; see also Watanabe \&
Yokoyama 2006;  Zenitani, Hesse \& Klimas 2009 for relativistic MHD
simulations that support this picture). 
High-$\sigma$ material is advected into the reconnection region where the release
of magnetic energy takes place.  Part of the dissipated magnetic energy 
serves to give bulk acceleration to a pair of minijets that move in
opposite directions (in the rest frame of the jet) and the
rest to accelerate particles in the minijets.
We explore the possibility that emission from minijets results in 
the observed flaring activity.

In this model, the material can leave the reconnection region with bulk
$\Gamma_{\rm co}$ close to the Alfv\'en speed of the upstream plasma:
$\Gamma_{\rm co}\sim \sqrt {\sigma}\simeq 10\sigma_2^{1/2}$ in the 
rest frame of the jet (Petschek 1964; Blackman 
\& Field 1994; Lyutikov \& Uzdensky 2003; Lyubarsky
2005). The energy density in the minijet is  (Lyubarsky 2005)
$\tilde{e}_{\rm em}\sim e'_{\rm j}=L_{\rm j, iso }/4\pi r^2c\Gamma_{\rm j}^2=0.05 L_{\rm
  j,45}r_2^{-2}\Gamma_{\rm j,5}^{-2}\quad \rm{erg/cm^3}.$

Even though we consider a Poynting flux-dominated jet, 
the (downstream) emitting region is not
necessarily magnetically dominated since a large part of the magnetic energy
dissipates in the reconnection region. This has important implications for the 
resulting spectra. Actually, Lyubarsky (2005) shows that material leaves
the reconnection  region with $\Gamma_{\rm co}\sim \sqrt {\sigma}$ for 
sufficiently small guide field. In this case, the magnetization of the 
minijet (downstream plasma) is $\sigma_{\rm em }\equiv \tilde{B}_{\rm em}^2/
4 \pi \tilde{e}_{\rm em}<1$, and the magnetic field in the minijet rest frame is parameterized as
$\tilde{B}_{\rm em}= \sqrt{\sigma_{\rm em}4\pi \tilde{e}_{\rm em}}=0.8  
\sigma_{\rm em}^{1/2}L_{\rm j,45}^{1/2}r_2^{-1}\Gamma_{j,5}^{-1}\quad
{\rm Gauss}$. 
If the guide field of the upstream plasma is stronger, the minijet is strongly magnetized 
with $\sigma_{\rm em}\simmore 1$, and slower, $\Gamma_{\rm co}< \sqrt
{\sigma}$. Furthermore, only a fraction $\sim 1/(1+\sigma_{\rm em})$
of the magnetic energy is dissipated in the reconnection region, leading
to a rather weaker and slower minijets. Here we focus on
fast minijets characterized by $\sigma_{\rm em}\simless 1$.

\subsection{Particle distribution}

Assuming an  electron-proton jet, the plasma coming into the reconnection
region (the upstream) contains
$\sim {\sigma} m_{\rm p}c^2$ magnetic energy per particle. Out of this energy
a fraction $\sqrt{\sigma}$ goes into bulk motions with the rest being
available to accelerate/heat particles. In the frame of the reconnection jet 
(minijet), the available energy per particle is $\sim \sqrt{\sigma} m_{\rm p}c^2$.
If electrons receive an appreciable fraction of the released magnetic 
energy $f\sim 0.5$, they are heated to characteristic thermal $\gamma$ factor
 \be \gamma_{\rm e, ch}\sim f \sqrt{\sigma}m_{\rm p}/m_{\rm e}\sim 10^4f_{1/2}
\sigma_2^{1/2}, \ee assumed to be isotropic in the blob rest frame.

In the context of the leptonic emission model discussed here, 
electrons of at least $\sim 10$ TeV (in the lab frame) 
are needed to potentially explain the observed emission that
extends up to $\sim 10$ TeV. This means that the random
component of the electron distribution should extend 
to $\gamma_{\rm e}>2\times 10^7/\Gamma_{\rm em}\gg \gamma_{\rm e, ch}$ (in the minijet
frame), where   $\Gamma_{\rm em}$ is the bulk Lorentz factor of the minijet.
It is thus clear that electrons need to be accelerated well above the
characteristic Lorentz factor $\gamma_{\rm e, ch}$. For simplicity, we will assume that the 
electrons follow a power-law distribution that extends 
to $\gamma_{\rm e}\gg \gamma_{\rm e, ch}$. It remains to be shown that relativistic
reconnection can lead to such a particle distribution. 

Since, for the parameters relevant to the jet of M87, particles with
$\gamma_{\rm e}\simmore
\gamma_{\rm e, ch}$ are fast cooling (because of synchrotron and, mainly, inverse
Compton emission; see next section), we consider emission coming from a rather steep 
particle distribution with $p\simmore 3$. The particle distribution has 
a maximum cutoff $\gamma_{\rm max}$. In our discussion,
$\gamma_{\rm max}$ is not important (provided that it is high enough to produce
 $\sim 10$ TeV photons through inverse Compton) and will be set to 
$\gamma_{\rm max}\to \infty$.  

\subsection{Radiation mechanisms}

The minijet contains energetic electrons that emit through synchrotron and  
synchrotron-self-Compton and, possibly, external inverse Compton (EIC)
mechanisms. We consider {\it any} source external to the minijet
contributing seed photons that are scattered by the minijet as an EIC source. 
Following this definition, not only disk emission but also soft photons 
originating from other parts of the jet
that interact with the minijet are labeled as EIC sources.      

In this Section, we explore the emission from a minijet for parameters
of the model relevant to M87 (i.e., we set $\theta=30^{\rm o}$, $\Gamma_{\rm j}=5$).
Other parameters of the model (i.e., magnetization of the jet, radius
of formation of the minijet) are less constrained. Since
we model M87 as a misaligned blazar, for illustration we set these parameters to values
inferred from modeling of Mrk 501 and PKS 2155-304 (Giannios et al. 2009). 
The Lorentz factor of the minijet in the rest frame of the jet is, therefore, set to
$\Gamma_{\rm co}=\sqrt{\sigma}=10$. For these parameters, using eqs. (1), (2) we find that the minijet 
moves with $\Gamma_{\rm em}=12$. The formation of the minijet is assumed to 
take place at $R_{\rm em}=100R_{\rm g}$ and the magnetization of the plasma
within the minijet is set to $\sigma_{\rm em}=1/3$ (though different values of
$\sigma$ and $\sigma_{\rm em}$ are
explored as well). From the observed luminosity ($L_{\rm f}\sim 10^{42}$
erg/sec) and duration ($T_{\rm f}\sim10^5$ sec) of the TeV flares, we can estimate
the lab-frame energy contained in the minijet $E_{\rm em}\sim L_{\rm
  f}T_{\rm f}/4\Gamma_{\rm em}^2 f\sim 5\times 10^{44}$ erg and the typical 
length scale of the emitting region is $l_{\rm em}\sim (E_{\rm
  em}/\Gamma_{\rm em}\tilde{e}_{\rm em})^{1/3}\sim 10^{15}$ cm.
The minijet can be easily powered by the M87 jet of $L_{\rm j,true}\sim 10^{43}$ erg/sec.
In the following, we show that the minijet emits its energy efficiently
in the $\sim$ TeV energy range. 

The synchrotron emission of electrons with random Lorentz factor $\gamma_{\rm e, ch}$ 
takes place at observed energy
$\nu_{\rm syn}\sim\Gamma_{\rm em}\gamma_e^2\nu_{\rm c}\simeq 5
L_{\rm j,45}^{1/2}\sigma_2^{3/2}f_{1/2}^2r_2^{-1}\quad {\rm eV}$ if the
minijet points at us, where $\nu_{\rm c}$ is the electron cyclotron frequency.
The self-Compton emission appears at 
$\nu_{\rm SSC}\sim \gamma_{\rm e}^2 \nu_{\rm syn}\simeq 0.5
L_{\rm j,45}^{1/2}\sigma_2^{5/2}f_{1/2}^4r_2^{-1}\quad {\rm GeV}$.  
Higher energy emission is expected by particles with $\gamma_{\rm
  e}>\gamma_{\rm e, min}$.

In addition to the synchrotron emission in the minijet,
ambient photon fields present in the nucleus of M87 are 
also a source of seed photons to be upscattered by the minijet. Some of these
soft photons come from the accretion flow (Di Matteo et al. 2003; Li et
al. 2009) and the rest from the inner jet. The observed nuclear emission has
a flat $Ef(E)$ spectrum in the range $\sim 0.01-10$ eV
with integrated luminosity $\sim 3\times 10^{41}$ erg/sec
declining at $E\simless 0.01$ eV and $E\simmore 10$ eV (e.g., Li et
  al. 2009). The nuclear emission has been resolved down to $\sim 20R_{\rm g}$
in the radio (see, e.g., Spencer \& Junor 1986; the radio emission is believed
to come from the jet). The resolution
is poorer in shorter wavelengths, corresponding to an upper limit
to the size of the nucleus of $\sim 100R_{\rm g}$ in the mm 
and $\sim 3\times 10^4 R_{\rm g}$ in the infrared (Whysong 
\& Antonucci 2004).

The {\it observed} emission from the nucleus competes with the synchrotron emission 
from the minijet as sources of seed photons to be inverse-Compton scattered in the minijet. 
A conservative estimate for the role of external soft photons 
comes from assuming that they are emitted isotropically from a 
region close to the black hole.
If, on the other hand, a large fraction of the observed soft photons comes from the jet
(i.e. they are emitted isotropically in the jet frame), 
their intensity along the jet axis is much larger than that estimated 
assuming isotropic emission from the inner disk. 

If the soft photon field is emitted isotropically from a region 
with length scale $R_{\rm soft}$, the typical angle $\theta_{\rm int}$ 
at which the soft photons interact with the bulk motion of the minijet depends 
on the ratio $R_{\rm soft}/R_{\rm em}$. For $R_{\rm soft}/R_{\rm em}\ll 1$,
the interaction angle equals approximately the inclination of the
minijet $\theta_{\rm int}\sim \theta\sim 30^{\rm o}$ while for $R_{\rm soft}/R_{\rm
  em}\simless 1$, the interaction angle is larger, $\theta_{\rm int}\sim
60^{\rm o}$. The energy density of the external photon field at the location of
the minijet is $\tilde{U}_{\rm ph}\sim (1-\beta_{\rm em}\cos\theta_{\rm int})
\Gamma_{\rm em}^2L_{\rm soft}/4\pi cR_{\rm
em}^2=0.04(1-\beta_{\rm em}\cos\theta_{\rm int})\Gamma_{\rm em,10}^2L_{\rm
  s,41.5}r_2^{-2}$ erg/cm$^3$ (in the rest frame of the
minijet), which appears to be of the same order of magnitude as that of the
magnetic field, $\tilde{U}_{\rm B}=8\times 10^{-3}L_{\rm j,45}
\sigma_{\rm em,1/3}r_2^{-2}\Gamma_{\rm j,5}^{-2}$ erg/cm$^3$. 
It should be noted that the
expressions for  $\tilde{U}_{\rm ph}$ and $\tilde{U}_{\rm B}$ hold for 
$\sigma_{\rm em}\simless 1$. For $\sigma_{\rm em}\simmore 1$ the magnetic
 energy density $\tilde{U}_{\rm B}$ saturates to a value that is
independent of $\sigma_{\rm em}$, while
for  $\sigma_{\rm em}\gg 1$, no minijet forms since the outflow from the 
reconnection region is expected to be slow and weak (since most of the 
magnetic energy is not dissipated
by reconnection).

For reasonable range of the parameters, either SSC or EIC can
dominate the production of the gamma rays.
First, we explore parameters for which synchrotron seed photons and SSC dominate
the bulk of the emission. We set $\sigma_{\rm em}\sim 1$ and assume a
very compact emission region for the disk photons: $R_{\rm soft}/R_{\rm em}\ll 
1$. The bulk Lorentz factor of the minijet is taken to be rather modest, $\Gamma_{\rm em}=8$
(by setting $\sigma=50$). Having specified the properties of the minijet and of the ambient
radiation field, we proceed with the calculation of the resulting
spectra. We calculate the synchrotron emission from a power-law distribution
of relativistic electrons using standard expressions.
The SSC and EIC emission are calculated using the
$\delta$-function approximation for single electron emission.
Klein-Nishina effects on the electron scattering are approximated by
using a step function for the energy dependence of the cross
section: $\sigma=\sigma_T$, for $\gamma_{\rm e} h\tilde \nu/m_{\rm e}c^2<3/4$
and $\sigma=0$ otherwise (see, for example  Tavecchio, Maraschi \& Ghisellini 1998 
and Coppi \& Blandford 1990 for discussion on the accuracy 
of this method).

\begin{figure}
\resizebox{\hsize}{!}{\includegraphics[angle=270]{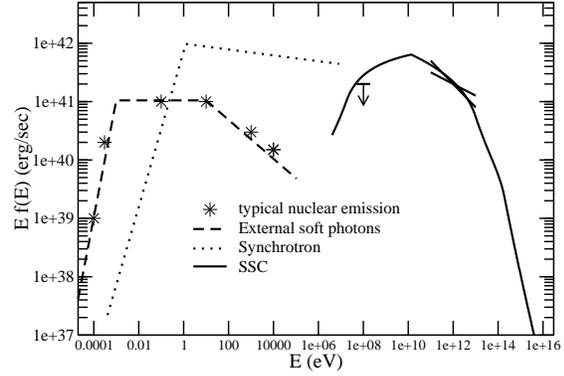}}
\caption[] {Synchrotron (dotted line) and inverse Compton (solid line) 
emission from a minijet forming at $R_{\rm em}=100 R_{\rm g}$ and beaming at an 
angle $\theta\sim 30^{\rm o}$ with respect to the axis of the jet of M87. 
The magnetization of the minijet is set to $\sigma_{\rm em}=1$
and a power-law electron distribution is assumed with index $p=3.1$.
The observed emission from the nucleus is sketched with the dashed line
and is assumed to come from a region $R_{\rm soft} \ll R_{\rm em}$.
The black bow-tie reports the TeV spectrum observed during the early 2008
flare (Albert et al. 2008). The EGRET 2$\sigma$ upper limit is also shown 
(Sreekumar et al. 1996).
      
\label{fig4}}
\end{figure}

The resulting spectra are shown in Fig.~4. Here we ignore any absorption
of the TeV emission due to pair creation (discussed in the next section).
The inverse Compton peak of the $Ef(E)$ spectrum $E_{\rm p}=\Gamma_{\rm
  em}\gamma_{\rm e,ch} m_{\rm e}c^2$
corresponds to the inverse Compton emission from electrons with 
the characteristic Lorentz factor $\gamma_{\rm e,ch}$. 
Since the peak is located below the working range of the Cherenkov 
telescopes, $E_{\rm p}\simless 100$ GeV (Aharonian et al. 2006; Albert et al. 2008), 
we have the constraint that $\Gamma_{\rm em}\gamma_{\rm e,ch}\simless 2\times
10^5$. Above the peak, the TeV spectrum is rather flat with photon-number
index $\Gamma\simeq -2.3$. The TeV spectrum depends on the index $p$ 
of the electron distribution (set to $p=3.1$).
The TeV emission becomes harder for lower $p$ and vice versa.   
At $E\sim 10^{13}$ eV the spectrum steepens. Photons observed at 
$E\simmore 10^{13}$ eV (i.e., $E\simmore 10^{12}$ eV in the rest frame of the
minijet) mainly come from scattering in the Thomson regime of optical synchrotron
photons (near-IR in the minijet frame),
the energy density of which drops rapidly with decreasing energy. 

The slope of the TeV emission spectrum 
is compatible with that observed during the fast TeV flaring in 
2005 (Aharonian et al. 2006) and 2008 (Albert et al. 2008).
At the same time the synchrotron emission is clearly pronounced
in the optical, UV and X-rays, far exceeding (by a factor of 10-30)
the typical nuclear emission. Note that, within the one-zone SSC model
for the TeV emission discussed here, the synchrotron component peaks
close to the optical and extends up to the hard X-rays rather independently
of the adopted parameters. 
For photons to be upscattered to energies $E\simmore 10$ TeV in the 
Thomson regime, the synchrotron emission cannot peak much above the 
optical wavelengths. Furthermore,
the electrons with random $\gamma_{\rm e}\sim 10^6$ needed to
produce the $E\simmore 10$ TeV emission emit into the hard X-ray regime
through synchrotron (for the magnetic field strength in the minijet 
of order 1 Gauss predicted by the model). Since powerful optical and X-ray 
flares with $\sim 1$ day timescales have not been observed in the nucleus despite the regular
monitoring of M87, we consider the SSC interpretation of the TeV 
emission unlikely.

\begin{figure}
\resizebox{\hsize}{!}{\includegraphics[angle=270]{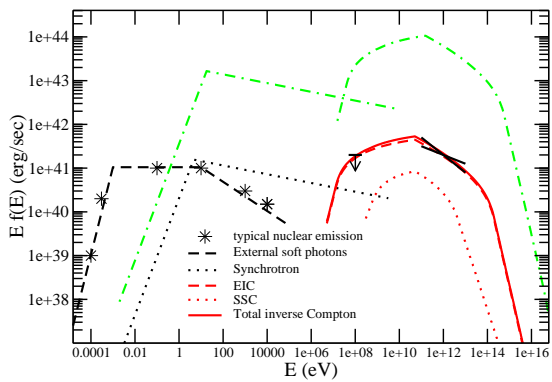}}
\caption[] {Same as in Fig.~4 but with
the magnetization of the minijet set to $\sigma_{\rm em}=1/3$
and the observed emission from the nucleus assumed to come from 
a larger region, $R_{\rm soft} \simless R_{\rm em}$.
A power-law electron distribution is assumed with index $p=3.2$.
For these parameter the EIC mechanism
dominates the gamma-ray emission. The green dash-dotted lines show the 
synchrotron and inverse Compton emission from an identical 
blob beaming within the jet axis as observed by an observer located
at an angle $\theta=1/\Gamma_{\rm j}=11^{\rm o}$ (with respect to the jet axis). 
      
\label{fig5}}
\end{figure}

The synchrotron component of the minijet is weaker when $\sigma_{\rm  em}<1$.
This brings us to the parameter space for which the EIC mechanism 
plays a more important role. We set $\sigma_{\rm em}=1/3$ (or
smaller) and assume an extended region for the external 
emissions: $R_{\rm soft}/R_{\rm em}\simless 1$. 
The bulk Lorentz factor is set to $\Gamma_{\rm em}=12$
(by setting $\sigma=100$).
The resulting spectra are shown in Fig.~5, where
 one can see that EIC dominates the GeV-TeV emission. For smaller
values of $\sigma_{\rm em}$ 
the synchrotron and SSC components become weaker but the TeV emission remains 
practically the same. The TeV spectrum is rather well decribed
with a power-law model with photon-number index $\Gamma\simeq -2.3$. 
The slope of the TeV spectrum depends on the spectrum
of the soft seed photons (observed to be $F_{\rm soft}(E)\propto E^{-1}$) 
and the index $p$ of the electron distribution (set to $p=3.2$).   
At $E\sim 3\times 10^{14}$ eV the spectrum steepens because 
the energy density of the available soft seed photons with
$E_{\rm s}\simless 10^{-3}$ eV for scattering in the Thomson regime  
drops rapidly with decreasing energy.  

The slope of the TeV emission spectrum 
is compatible with that observed during the fast TeV flaring in 
2005 (Aharonian et al. 2005) and 2008 (Albert et al. 2008).
During the TeV flares the $\sim 100$ MeV flux approaches
the upper limits set by {\it EGRET} (see Fig.~5). If M87 is currently in a high
GeV-TeV state, we predict that {\it FERMI} should soon have a significant detection
of the nuclear emission.  

During flares, the synchrotron emission from the minijet dominates
the nuclear emission in the X-rays, provided that $\sigma_{\rm em}\simmore 0.1$. In this case,
simultaneous X-ray flares of moderate strength are expected. 
Unfortunately, there have been no X-ray
observations strictly simultaneous with the TeV flares. On the other hand,
monitoring of the nucleus in X-rays in 2008 revealed that it was in a high
X-ray state around the time of the TeV flares. Simultaneous multi-wavelength
observations can greatly constrain magnetization of the emitting region.

In Fig.~5, we also show for comparison the emission expected from M87
if it were observed at an inclination $\theta=1/\Gamma_{\rm j}\sim 11^{\rm o}$, i.e., 
at $\alpha=1$. In that case, 
the flares would be more powerful, powered by minijets emitting within the jet axis,
and have observed durations of $\sim$several hours. 
The synchrotron emission would peak in the extreme UV and the inverse Compton
component at $\sim 100$ GeV. Both the timing and
spectral properties would look similar to those of a low luminosity, high-frequency 
BL Lac (HBL) object. Note, however, that in HBLs the inverse Compton peak
is typically not as pronounced with respect to the synchrotron one as 
that of Fig.~5.       

\subsubsection{Escape of the TeV photons}

The spectrum of the TeV emission from M87 does not show any signature 
of absorption up to $\sim 10$ TeV. In this section,
we examine the conditions under which the TeV emission from a misaligned
minijet escapes pair creation in the region where it is produced. 

The dominant source of infrared-optical photons that can pair create with the
TeV photons is the ambient nuclear emission\footnote{The synchrotron photons from
the minijet may, in principle, also contribute to the TeV absorption. In
practice, however, the strong beaming of the minijets toward the observer
guarantees that the synchrotron photons can be ignored as a source of opacity.}. 
The size of the emitting region (which likely depends on
the wavelength of the emission) and the location and direction of the minijet
determine the typical angle $\theta_{\rm int}$ at which the TeV photons interact
with soft photons. For a rough estimate of this angle we set 
$R_{\rm soft}\simless R_{\rm em}$, resulting in $\theta_{\rm int}\sim 60^{\rm
  o}$. The pair-creation cross section peaks at 
$\sigma_{\gamma\gamma}\sim \sigma_T/5$ for photons with total
energy $E_{\rm CM}\sim 4 m_{\rm e}c^2$ in the center of mass frame. The center-of-mass 
energy of a hard photon $E_{\rm h}$ and a soft photon $E_{\rm s}$ that form 
an angle $\theta_{\rm int}\sim 60^{\rm o}$ is 
$E_{\rm CM}=\sqrt{2E_{\rm h}E_{\rm s}(1-\cos\theta)}\sim \sqrt{E_{\rm
h}E_{\rm s}}$. The $\gamma \gamma$ annihilation cross section peaks at soft photon energy
$E_{\rm s}\simeq 4/E_{\rm h,1TeV}$ eV. Since the number density of soft photons
increases at the lower energies, $\sim$10 TeV photons are the most likely to
be absorbed by $\sim 0.4$ eV target soft photons.
The number density of soft photons is $N_{\rm target}=L_{\rm IR}/4\pi R_{\rm em}^2
cE_{\rm s}$, where $L_{\rm IR}\sim 10^{41}$ erg/sec is the near-infrared luminosity of the
nucleus (see Figs.~4, 5). Thus, the annihilation optical depth is 
\be
\tau_{\gamma\gamma}=
N_{\rm target}\sigma_{\gamma\gamma}R_{\rm em}\simeq 1L_{\rm IR,41}E_{\rm
  h,10TeV}/r_2.
\ee
From this expression we conclude that for $R_{\rm soft}\simless R_{\rm em}$,
the $\sim 10$ TeV photons marginally  avoid
significant attenuation if the minijet forms at $R_{\rm em}\sim 100R_{\rm g}$.   
If, however, $R_{\rm soft}$ is much smaller than $R_{\rm em}$ then escape is possible from smaller radii.

\section{Discussion/Conclusions}  

M87 is a well studied case of a misaligned AGN jet. Both the nucleus
of M87 and the knot HST-1 have variable emission on timescales
of weeks to months, in particular in the X-rays (e.g., Harris et al. 2009).
More impressively, M87 reveals compact TeV emitting regions by producing flares on $\sim 1$ day
timescales (Aharonian et al. 2006; Albert et al. 2008).
VLBI observations during and after the 2008 TeV flares showed the ejection
of two blobs within a distance of $\sim 100R_{\rm g}$ from the black hole,
strongly suggesting the nucleus as the source of the flare (Acciari et al. 2009).

The TeV flares may originate from the magnetosphere of the black hole
(Levinson 2000; Neronov \& Aharonian 2007) or a complex jet geometry 
(spine/layer interaction [Tavecchio \& Ghisellini 2008]; 
collimating jet [Lenain et al. 2008]; decelerating jet [Georganopoulos et al. 2005]).
Still, in models in which the jet is misaligned with respect to our line of sight,
the Doppler factor of the emitting material is $\delta\sim 1$, 
implying the need to explain a very compact
emitting region of length scale $l_{\rm em}\simless 5 R_{\rm g}$
and the lack of TeV absorption.

Rapid gamma-ray flaring has been observed frequently in blazars. 
Mrk 501 and PKS 2155-304 show powerful flares
at TeV energies with duration of several minutes,
much shorter than the light crossing time of their central 
black holes (Aharonian 2007; Albert et al. 2007). 
This extremely fast flaring suggests that the flares do not
reflect variability of the central engine in these jets but
is connected to interactions within the jet, which lead to compact
emitting regions and potentially relativistic motions in the rest frame 
of the jet. 

In this line of argument, we have previously proposed a 
jets-in-jet model for blazars  (Giannios et al. 2009).  
Minijets, driven by magnetic reconnection and moving
relativistically within the main jet, can power the fast evolving flares, 
make them transparent at TeV energies, 
and explain why the jet as a whole appears much slower.
The reconnection may be triggered by MHD instabilities in the jet
(like kinks) and/or reversals of the polarity of the
magnetic field in the magnetosphere of the black hole or in the inner disk.
When a field reversal takes place, interactions/collisions 
of parts of the jet with opposite
polarity can lead to very efficient release of magnetic energy via magnetic reconnection.

Reconnection within the jet produces two oppositely directed
(in the jet's frame) minijets. One of them always points within
the jet angle and is observable in blazars because their jets point at us.
The other minijet (its counterpart) points outside the jet opening angle
and is potentially observable to off-axis observers in case of misaligned
jets, e.g., M87 or Cen A (Aharonian et al. 2009).

The TeV flaring of M87 can be understood in this context.
Furthermore, we predict that if M87 is currently in a high
GeV-TeV state, {\it FERMI} should soon have a significant detection
of the nuclear emission. Finally, if viewed on-axis, M87 would look like a fairly 
regular high peaked BL-Lac object with TeV (and possibly optical/Xray) flares of several hours. 

The high-energy emission from the minijets comes during their formation/ejection
and/or the (short) cooling timescale associated with the TeV emitting
particles. The observed TeV flares are likely part of a general increase of 
minijet events in the jet, most of which we do not see because 
they are beaming away from us. On a longer timescale, the  minijets interact
with the rest of the jet and are slowed down (with respect to the jet).
During this interaction phase, shocks are expected to form, resulting 
in further acceleration of particles that can power an `afterglow'
emission. Interestingly the Acciari et al. (2009) VLBI observations
that followed the February 2008 flaring of M87 showed the ejection
of radio blobs from the nuclear region. It is tempting to 
associate these blobs with the collective afterglows from minijets 
brought to our view after their deceleration to the jet mean motion.

By analogy with the $\sim 5$ min flares seen from  Mrk 501 and PKS 2155-304, 
one might expect even faster evolving flares from M87 than the observed 
$\sim 1$ day ones. Taking into account 
that the black hole of M87 is a factor of $\sim 3$ larger than those
of Mrk 501 and PKS 2155-304 and that the large
inclination of the M87 jet dilutes timescales by a factor of $\alpha^2\sim 10$, 
one could still expect flares evolving in as short as $2-3$ hours from M87.

\section*{Acknowledgments}
DG acknowledges support from the Lyman Spitzer Jr. Fellowship awarded by the
Department of Astrophysical Sciences at Princeton University.
DG acknowledges the support of the NORDITA program on 
{\it Physics of relativistic flows} during which part of this work
was completed. DU is supported by National Science Foundation Grant 
No.\, PHY-0821899 (PFC: Center for Magnetic Self-Organization 
in Laboratory and Astrophysical Plasmas). MCB acknowledges support from NASA, 
via a {\it Fermi Gamma-ray Space Telescope} Guest Investigator grant.

\end{document}